# Electronic Fingerprints of Cr and V Dopants in Topological Insulator Sb$_2$Te$_3$

Wenhan Zhang[1], Damien West[2], Seng Huat Lee[3,4], Yunsheng Qiu[3], Cui-Zu Chang[5,6], Jagadeesh S. Moodera[6], Yew San Hor[3], Shengbai Zhang[2] and Weida Wu[1, *]

[1]Department of Physics and Astronomy, Rutgers University, Piscataway, New Jersey 08854, United States

[2]Department of Physics, Applied Physics, and Astronomy, Rensselaer Polytechnic Institute, Troy, NY 12180-3590, United States

[3]Department of Physics, Missouri University of Science and Technology, Rolla, Missouri 65409, United States

[4]2D Crystal Consortium, Materials Research Institute, The Pennsylvania State University, University Park, Pennsylvania 16802, United States.

[5]Department of Physics, The Penn State University, University Park, PA 16802, United States

[6]Francis Bitter Magnet Lab, Massachusetts Institute of Technology, Cambridge, Massachusetts 02139, United States

*Correspondence should be addressed to wdwu@physics.rutgers.edu.

**Abstract: By combining scanning tunneling microscopy/spectroscopy and first-principles calculations, we systematically study the local electronic states of magnetic dopants V and Cr in the topological insulator (TI) $Sb_2Te_3$. Spectroscopic imaging shows diverse local defect states between Cr and V, which agree with our first-principle calculations. The unique spectroscopic features of V and Cr dopants provide electronic fingerprints for the co-doped magnetic TI samples with the enhanced quantum anomalous Hall effect. Our results also facilitate the exploration of the underlying mechanism of the enhanced quantum anomalous Hall temperature in Cr/V co-doped TIs.**

## I. INTRODUCTION

Quantum anomalous Hall (QAH) state offers a promising platform to develop novel low energy-consuming electronic and spintronic devices and hence attracts a wide research interest in condensed matter physics. It is characterized by the spin-polarized dissipationless chiral edge states, which is a result of the topologically nontrivial band structures with the broken time-reversal symmetry [1]. The time-reversal symmetry can be broken by introducing ferromagnetic (FM) order, via either the magnetic doping [1] or the magnetic proximity effect in topological insulator (TI)/FM heterostructures [2]. To date, the chromium (Cr)- and vanadium (V)-doped $(Bi,Sb)_2Te_3$ are the only two systems that harbor the QAH state [3,4]. However, the full quantization in both systems requires extremely low temperature (T< 100 mK). Recently, Ou *et al.* achieved a significant increase of the characteristic QAH temperature in Cr/V co-doped $(Bi,Sb)_2Te_3$ films [5], where the critical temperature enhancement has been attributed to the improvement of the FM order [3,4,6–9].

FM order has been reported in TIs doped with a number of 3d transition metal elements, e.g. Cr [10–12], V [13–15] and Mn [16,17]. However, there is no consensus on the mechanism of the FM ordering in magnetically doped TIs. The carrier-mediated exchange interaction (i.e. Ruderman-Kittel-Kasuya-Yosida interaction) has been predicted to be responsible for FM order especially in the presence of free bulk carriers in TIs [13,18], similar to the case in conventional diluted magnetic semiconductors [19–22]. In addition, a carrier-independent exchange interaction known as van Vleck mechanism was also proposed in magnetically doped TIs, which is the basis for the realization of QAH state [1,23]. The van Vleck mechanism is independent of the spatial distribution of magnetic dopants. More recently, first-principles calculation studies suggest a new carrier-independent exchange interaction in magnetically doped TI [24,25]. In this new mechanism, the 3*d* orbitals of magnetic dopants hybridize with the surrounding *p*-bonding network, resulting in a net spin density localized around magnetic dopants. Thus, the overlap of the hybridized states mediates the exchange coupling between magnetic dopants. Therefore, it is imperative to visualize the spatial distribution of these hybridized 3*d* states of magnetic dopants and to study the underlying exchange mechanism. Scanning tunneling microscopy/spectroscopy (STM/STS) is a powerful technique to image local electronic states with atomic resolution. So far, there are a few of STM studies on Cr- and V doped TIs [26–29], only two of which are about the spatial distribution of the magnetic defect states [30,31]. The comparison of the spatial distribution of the magnetic defect states between two parent systems for the QAH state (i.e. Cr- and V- doped $Sb_2Te_3$) is still lacking. Such a study will also facilitate the understanding of the enhancement of QAHE critical temperature in Cr/V co-doped $(Bi,Sb)_2Te_3$ [5,32].

In this report, we present systematic STM/STS and first-principle calculation studies on local electronic defect states of V and Cr dopants in $Sb_2Te_3$. Our work shows that the defect states of Cr are well above the conduction band minimum (CBM), while the defect states of V are located near the valence band maximum (VBM), consistent with density functional theory (DFT) calculations. Our results demonstrate that the defect states of Cr and V dopants can serve as electronic fingerprints to distinguish them in TIs. This study will pave the way to investigate the mechanism of enhanced FM order in Cr/V co-doped TIs and singling out materials candidates for the high-temperature QAH state.

## II. EXPERIMENTAL AND CALCULATION DETAILS

STM/STS measurements were performed at 4.5 K in an Omicron UHV-LT-STM with a base pressure $1 \times 10^{-11}$ mbar. Electrochemically etched tungsten tips were characterized on single crystal Au (111) surface. The V- and Cr-doped $Sb_2Te_3$ crystals were cleaved *in-situ* at room temperature and then transferred immediately into the cold STM head for measurements. The d$I$/d$V$ spectra mapping measurements were performed with the standard lock-in technique with a modulation frequency $f$ = 455 Hz and a modulation amplitude $V_{mod}$ = 20 mV.

Our first-principles calculations are based on DFT within the approximation of Perdew-Burke-Ernzerhof [33]. Interactions between ion cores and valence electrons are described by the projector augmented-wave (PAW) method [34] as implemented in the VASP package [35,36]. Plane waves with a kinetic energy of less than 266 eV were used as the basis set. The spin-orbit interactions were implemented in the all-electron part of the PAW Hamiltonian within the muffin tin spheres. Simulated STM images were constructed using the theory of Tersoff and Hamann [37] wherein the charge density was determined self-consistently with Gamma point sampling of the Brillouin zone. The supercells used for calculations contained a quintuple layer of $Sb_2Te_3$ with lateral dimensions of (8×8) containing 320 atoms or three quintuple layers with lateral dimensions (4×4) containing 240 atoms, for simulated STM and band structure calculations, respectively. A minimum of 10 Å of vacuum was used to separate slabs from their periodic images. Self-consistent charge densities associated with the calculation of the band structures were obtained using a 3×3×1 Monkhorst-Pack mesh k-point sampling [38]. In the determination of the defect geometries, all atoms were allowed to relax until the maximum Hellmann-Feynman force was converged to within 0.02 eV $Å^{-1}$.

The single crystals of V-doped and Cr-doped $Sb_2Te_3$ were grown by modified Bridgman methods. The stoichiometric mixtures of Sb (99.999%), Te (99.999%) and V (99.7%) or Cr (99.99%) with various nominal $x$ values were heated up to 950 ℃ at the rate of 1 ℃/min in evacuated quartz tubes. The melts were maintained at the temperature for two days. Then they were cooled to 550 ℃ at the rate of 0.1 ℃/min. The crystalline ingots were then quenched in cold water after a day of the 550 ℃ annealing.

The V-doped $Sb_2Te_3$ films were grown by a custom-built ultrahigh vacuum molecular beam epitaxy (MBE) system [3]. 10 nm Te was deposited on the top as a capping layer for *ex-situ* STM measurements. The Te capping layer was removed by sputtering and annealing the samples before STM measurements.

## III. RESULTS AND DISCUSSION

Figs. 1(a-b) show representative topographic images of cleaved surfaces of V- and Cr-doped $Sb_2Te_3$ crystals. No interstitial V or Cr atoms were observed. Presumably, all the V and Cr dopants occupy Sb sites, forming $V_{Sb}$ and $Cr_{Sb}$ defects at 2B or 4A sites (boxed in Figs. 1(a-b)) in the topmost quintuple layer (QL) as illustrated in Fig. 1(c). With sub-unit cell resolution, these magnetic defects at different depths can be unambiguously identified. The color images in Fig. 1(d) show the individual Cr and V defects. The magnetic defects at 2B sites are characterized by small dark triangles, while those at 4A sites are characterized by larger triangles with depressed corners and centers. The defects at 2B and 4A sites are crystallographically equivalent in the bulk, and thus the different appearances result from their different depths from the surface. Another type of prominent defects observed in both V- and Cr- doped $Sb_2Te_3$ is the native $Sb_{Te}$ anti-site, which has been reported in the prior work [39]. The $Sb_{Te}$ anti-site defects are marked by blue circles in Figs. 1(a-b) (see Appendix A).

The grey-scale images in Fig. 1(d) are the DFT simulated topographic images of V and Cr at 2B and 4A sites. The excellent agreement between theory and experimental results further substantiates the identification of the magnetic defects. By counting V and Cr defects within the topmost QL, their concentrations can be quantitatively estimated: V (Cr) takes up 0.25% (0.4%) of each Sb layer. Because V and Cr dopants appear almost identical in topography at the typical sample bias (e.g., −1 V), it would be difficult to distinguish them in the Cr/V co-doped QAH films. Therefore, it is crucial to establish reliable criteria for differentiating V and Cr dopants. To address this issue, we perform bias-dependent topography measurements of the two materials. Fig. 2 shows the comparison of STM images of individual V and Cr defects at different sample biases. Interestingly, Cr (2B) appears as a triangular depression. It does not change much except at $V=$ +0.6 V, where its shape rotates by 180 °. In contrast, we observed two types of V defects with different appearances. They are denoted as $V_I$ and $V_{II}$, respectively. Note that both of them occupy 2B or 4A sites, i.e., substituting Sb atoms. At high sample bias ($|V| \geq 0.6$ V), their appearances are almost identical to that of Cr. On the other hand, the two types of V and Cr can be distinguished at low bias ($|V| < 0.6$ V). $V_I$ (2B) becomes a protrusion surrounded by 3 petals at −0.2 V (the first row in Fig. 2), while $V_{II}$ (2B) has a similar appearance at +0.2 V (the third row in Fig. 2). The changes of defect appearances of Cr and V at 4A sites are similar to those at 2B sites. The bias-dependent topographic images shown in Fig. 2 were also observed in V-doped TI films, as shown in Fig. 3. The dramatic changes of the defect appearance, *i.e.*, $V_I$ at −0.2 V, $V_{II}$ at +0.2 V and Cr at +0.6 V, suggest that V and Cr induce strong modulations in the local density of states (LDOS) in $Sb_2Te_3$ at different energies. Therefore, to examine the local electronic defect states of V and Cr, we systematically measured the differential conductance (d$I$/d$V$), which is proportional to the LDOS.

Fig. 4(a) shows the average d$I$/d$V$ spectra measured in selected regions of V-doped $Sb_2Te_3$. The blue spectrum was measured in defect-free areas. The steep increases in the d$I$/d$V$ spectrum below $E_F$ and above $E_F$ + 0.3 V indicate the positions of the VBM and the CBM. The inset shows a V-shaped d$I$/d$V$ spectrum due to the linear dispersion of the topological surface states. The minimum indicates that the Dirac point ($E_D$) is at 0.12 eV above $E_F$, in agreement with previous studies [39,40]. Interestingly consistent with the topographic images shown in Fig. 2, the two types of V have different d$I$/d$V$ spectra. $V_I$ has two peaks associated with the defect states: one at $E_D -$ 0.20 eV, below the VBM; the other at $E_D +$ 0.05 eV, inside the band gap. As for $V_{II}$, a pronounced d$I$/d$V$ peak is located at $E_D -$ 0.07 eV, which is also in the band gap. The different defect states of the two types of V are the electronic origin of the dramatic change of their topographic appearances at low sample bias.

The spatial distributions of the defect states of both $V_I$ and $V_{II}$ at 2B and 4A sites were visualized by spectroscopic imaging. Figs. 4(d-f) show the d$I$/d$V$ maps taken at $E_D -$ 0.20 eV and $E_D +$ 0.05 eV, where the defect states of $V_I$ emerge. The defect states of $V_I$ (2B) at $E_D -$ 0.20 eV show a clover-like shape with three petals extending over ~2.1 nm. The in-gap defect states of $V_I$ (2B) are clearly visible at $E_D +$ 0.05 eV, which are spatially more extended (~3.2 nm). On the other hand, Fig. 4(e) show that the defect states of $V_{II}$ (2B) stand out at $E_D -$ 0.07 eV, corresponding to the in-gap peak in the d$I$/d$V$ spectrum. As for the V dopants occupying 4A sites, the defect states of $V_{II}$ (4A) at $E_D -$ 0.20 eV are invisible. They are probably smeared out by the bulk states of $Sb_2Te_3$ since $V_{II}$ (4A) is deeper from the cleaved surface than $V_{II}$ (2B). The intensity of in-gap defect states of $V_{II}$ (4A) at $E_D +$ 0.05 eV and $V_I$ (4A) at $E_D -$ 0.07 eV is weaker and more extended (~4.1 nm) than V (2B). The appearances of the in-gap defect states of $V_I$ and $V_{II}$ are very similar, suggesting a similar electronic origin.

The crystal field in the octahedral environment of V dopants splits the $d$-levels into triplet $t_{2g}$ and doublet $e_g$ states as shown in Fig. 1(e). The majority spin channel of the low lying $t_{2g}$ states is occupied with 2 electrons leading to magnetic moments of 2 $\mu_B$ per vanadium ion. The band structure of $V_{0.03}Sb_{1.97}Te_3$ calculated via DFT is shown in Fig. 4(b). The calculated partial density of states (pDOS) of V shows a peak at $E_D$, which is attributed to the impurity bands inside the band gap mainly coming from V $t_{2g}$ orbitals. This feature is in better agreement with $V_{II}$ defect states. Consistently, the spatial distribution of defect states in the DFT simulation (grey-scale images in Fig. 4(g)) resembles the clover-leaf shape as observed in experiment (color images in Fig. 4(g)).

Although a clear difference between the defect states of $V_I$ and $V_{II}$ was captured, the origin remains unclear. One possible scenario is that a slight Jahn-Teller distortion occurs around $V_I$ dopants and further splits their $t_{2g}$ orbitals. However, our STM is not able to resolve any local distortion around $V_I$ dopants. Alternatively, it may originate from distinct electronic environments induced by other charge defects. We did not find a clear correlation between V dopants and other charge defects (e.g. $Sb_{Te}$). The physical origin of $V_I$ and its impact on exchange coupling will be investigated in future studies. Similar spatial distribution of the V defect states was also observed in V-doped $Sb_2Te_3$ films shown in Fig. 5, supporting that the observed V defect states are intrinsic in $Sb_2Te_3$. If the concentration of V dopants is sufficiently high, these local defect states would overlap and likely form impurity bands. Recent ARPES studies suggest that V doping causes the VBM of V-doped $(Bi,Sb)_2Te_3$ to lie higher in energy than the Dirac point [18]. In light of our observations of pronounced local defect states near the Dirac point, the upshift of the VBM probably originates from the formation of the impurity band from V defect states.

Fig. 6(a) shows the $dI/dV$ spectra taken on Cr defect sites (red) and defect-free areas (blue). Cr was not observed to contribute defect states inside the band gap. Surprisingly, a peak in the Cr $dI/dV$ spectrum emerges at $E_D$ + 0.48 eV, which is well above the CBM. Fig. 6(c) shows a representative topographic image with atomic resolution where individual Cr dopants at 2B and 4A sites were observed. Fig. 6(d) shows the spatial distribution of Cr defect states visualized with $dI/dV$ imaging at $E_D$ + 0.48 eV. As far as we are aware of, these Cr defect states have not been reported in prior studies. To confirm the experimental observation, we calculated the band structure and the Cr pDOS in $Cr_{0.03}Sb_{1.97}Te_3$ via DFT. The results are shown in Fig. 6(b). The hybridized band with dominant $e_g$ orbitals of Cr is deep inside the conduction band, in good agreement with the experimental results. Furthermore, the calculated spatial distribution of Cr defect states (Fig. 6(e)) is consistent with experimental observation. The excellent agreement between theory and experiment provides compelling evidence that the observed Cr defect states are intrinsic in TIs.

## IV. CONCLUSION

In summary, our spectroscopic imaging shows that the V defect states contribute finite LDOS in the band gap of TIs because of the partially filled $t_{2g}$ orbitals. In contrast, the Cr defect states are deep in the conduction bands. Both V and Cr defects show nontrivial spatial distributions of local electronic defect states, which is corroborated by the DFT calculations. The unique features of V and Cr defect states can be used as electronic fingerprints to investigate the interaction between dopants in Cr/V co-doped TIs with higher temperature QAH state. The spectroscopic imaging studies of co-doped TIs may shed new light on the magnetic exchange mechanism and could lead to the QAH state realized at elevated temperatures for potential technological applications.

## ACKNOWLEDGMENTS

The STM studies at Rutgers were supported by NSF (Grant No. DMR-1506618). S. B. Z. was supported by the Department of Energy under Grant No. DE-SC0002623. D. W. was supported by NSF under Grant No. EFMA-1542789. The supercomputer time was provided by the CCNI at RPI and NERSC under DOE Contract No. DE-AC02-05CH11231. The single crystal synthesis works were supported by NSF under Grant No. DMR-1255607. C.Z.C was supported by Alfred P. Sloan Research Fellowship and the Army Research Office under Agreement No. W911NF1810198.

## APPENDIX A: NATIVE DEFECTS IN $Sb_2Te_3$

$Sb_{Te}$ antisites are common native defects in $Sb_2Te_3$. They have been observed in both V- and Cr-doped $Sb_2Te_3$. In Fig. 7, Two shapes of such defects (blue circles) are indeed visible: one is a bright protrusion, the other is a subtle triangle with a dark center. The topographic images with atomic resolution in Fig. 7(a) indicates that they are at the 1A and 5B sites, respectively. Their densities are statistically the same, consistent with their crystallographical equivalence. The $Sb_{Te}$ at the 3C site is not observed on the two materials. This may be because $Sb_{Te}$ at 3C has higher formation energy than $Sb_{Te}$ at 1A or 5B.

## REFERENCES

[1]    R. Yu, W. Zhang, H.-J. Zhang, S.-C. Zhang, X. Dai, and Z. Fang, Science. **329**, 61 (2010).

[2]    X.-L. Qi, T. L. Hughes, and S.-C. Zhang, Phys. Rev. B **78**, 195424 (2008).

[3]    C.-Z. Chang, W. Zhao, D. Y. Kim, H. Zhang, B. A. Assaf, D. Heiman, S.-C. Zhang, C. Liu, M. H. W. Chan, and J. S. Moodera, Nat. Mater. **14**, 473 (2015).

[4]    C.-Z. Chang, J. Zhang, X. Feng, J. Shen, Z. Zhang, M. Guo, K. Li, Y. Ou, P. Wei, L.-L. Wang, Z.-Q. Ji, Y. Feng, S. Ji, X. Chen, J. Jia, X. Dai, Z. Fang, S.-C. Zhang, K. He, Y. Wang, L. Lu, X.-C. Ma, and Q.-K. Xue, Science. **340**, 167 (2013).

[5]    Y. Ou, C. Liu, G. Jiang, Y. Feng, D. Zhao, W. Wu, X.-X. Wang, W. Li, C. Song, L.-L. Wang, W. Wang, W. Wu, Y. Wang, K. He, X.-C. Ma, and Q.-K. Xue, Adv. Mater. **30**, 1703062 (2018).

[6]    J. G. Checkelsky, R. Yoshimi, A. Tsukazaki, K. S. Takahashi, Y. Kozuka, J. Falson, M. Kawasaki, and Y. Tokura, Nat. Phys. **10**, 731 (2014).

[7]    X. Kou, S.-T. Guo, Y. Fan, L. Pan, M. Lang, Y. Jiang, Q. Shao, T. Nie, K. Murata, J. Tang, Y. Wang, L. He, T.-K. Lee, W.-L. Lee, and K. L. Wang, Phys. Rev. Lett. **113**, 137201 (2014).

[8]    Z. Zhang, Y.-H. Wang, Q. Song, C. Liu, R. Peng, K. A. Moler, D. Feng, and Y. Wang, Sci. Bull. **60**, 1301 (2015).

[9]    A. J. Bestwick, E. J. Fox, X. Kou, L. Pan, K. L. Wang, and D. Goldhaber-Gordon, Phys. Rev. Lett. **114**, 187201 (2015).

[10]   Z. Zhou, Y.-J. Chien, and C. Uher, Phys. Rev. B **74**, 224418 (2006).

[11]   J. S. Dyck, Č. Drašar, P. Lošt'ák, and C. Uher, Phys. Rev. B **71**, 115214 (2005).

[12]   V. A. Kulbachinskii, P. M. Tarasov, and E. Brück, J. Exp. Theor. Phys. **101**, 528 (2005).

[13]   J. S. Dyck, P. Hájek, P. Lošt'ák, and C. Uher, Phys. Rev. B **65**, 115212 (2002).

[14]   J. S. Dyck, P. Švanda, P. Lošt'ák, J. Horák, W. Chen, and C. Uher, J. Appl. Phys. **94**, 7631 (2003).

[15]   Z. Zhou, Y.-J. Chien, and C. Uher, Appl. Phys. Lett. **87**, 112503 (2005).

[16]   Y. S. Hor, P. Roushan, H. Beidenkopf, J. Seo, D. Qu, J. G. Checkelsky, L. A. Wray, D. Hsieh, Y. Xia, S.-Y. Xu, D. Qian, M. Z. Hasan, N. P. Ong, A. Yazdani, and R. J. Cava, Phys. Rev. B **81**, 195203 (2010).

[17]   J. Choi, S. Choi, J. Choi, Y. Park, H.-M. Park, H.-W. Lee, B.-C. Woo, and S. Cho, Phys. Status Solidi **241**, 1541 (2004).

[18]   W. Li, M. Claassen, C.-Z. Chang, B. Moritz, T. Jia, C. Zhang, S. Rebec, J. J. Lee, M. Hashimoto, D.-H. Lu, R. G. Moore, J. S. Moodera, T. P. Devereaux, and Z.-X. Shen, Sci.

Rep. **6**, 32732 (2016).

[19]  T. Story, R. R. Gałązka, R. B. Frankel, and P. A. Wolff, Phys. Rev. Lett. **56**, 777 (1986).

[20]  H. Ohno, J. Magn. Magn. Mater. **200**, 110 (1999).

[21]  H. Ohno, D. Chiba, F. Matsukura, T. Omiya, E. Abe, T. Dietl, Y. Ohno, and K. Ohtani, Nature **408**, 944 (2000).

[22]  T. Dietl, Nat. Mater. **9**, 965 (2010).

[23]  M. Li, C.-Z. Chang, L. Wu, J. Tao, W. Zhao, M. H. W. Chan, J. S. Moodera, J. Li, and Y. Zhu, Phys. Rev. Lett. **114**, 146802 (2015).

[24]  J. Kim, S.-H. Jhi, A. H. MacDonald, and R. Wu, Phys. Rev. B **96**, 140410 (2017).

[25]  C. Chan, X. Zhang, Y. Zhang, K. Tse, B. Deng, J. Zhang, and J. Zhu, Chinese Phys. Lett. **35**, 017502 (2018).

[26]  F. Yang, Y. R. Song, H. Li, K. F. Zhang, X. Yao, C. Liu, D. Qian, C. L. Gao, and J.-F. Jia, Phys. Rev. Lett. **111**, 176802 (2013).

[27]  C.-Z. Chang, P. Tang, Y.-L. Wang, X. Feng, K. Li, Z. Zhang, Y. Wang, L.-L. Wang, X. Chen, C. Liu, W. Duan, K. He, X.-C. Ma, and Q.-K. Xue, Phys. Rev. Lett. **112**, 056801 (2014).

[28]  I. Lee, C. K. Kim, J. Lee, S. J. L. Billinge, R. Zhong, J. A. Schneeloch, T. Liu, T. Valla, J. M. Tranquada, G. Gu, and J. C. S. Davis, Proc. Natl. Acad. Sci. **112**, 1316 (2015).

[29]  Y. Jiang, C. Song, Z. Li, M. Chen, R. L. Greene, K. He, L. Wang, X. Chen, X. Ma, and Q.-K. Xue, Phys. Rev. B **92**, 195418 (2015).

[30]  P. Sessi, R. R. Biswas, T. Bathon, O. Storz, S. Wilfert, A. Barla, K. A. Kokh, O. E. Tereshchenko, K. Fauth, M. Bode, and A. V. Balatsky, Nat. Commun. **7**, 12027 (2016).

[31]  M. F. Islam, C. M. Canali, A. Pertsova, A. Balatsky, S. K. Mahatha, C. Carbone, A. Barla,

K. A. Kokh, O. E. Tereshchenko, E. Jiménez, N. B. Brookes, P. Gargiani, M. Valvidares, S. Schatz, T. R. F. Peixoto, H. Bentmann, F. Reinert, J. Jung, T. Bathon, K. Fauth, M. Bode, and P. Sessi, Phys. Rev. B **97**, 155429 (2018).

[32]  W. Wang, Y. Ou, C. Liu, Y. Wang, K. He, Q.-K. Xue, and W. Wu, Nat. Phys. (2018).

[33]  J. P. Perdew, K. Burke, and M. Ernzerhof, Phys. Rev. Lett. **77**, 3865 (1996).

[34]  P. E. Blöchl, Phys. Rev. B **50**, 17953 (1994).

[35]  G. Kresse and J. Furthmüller, Phys. Rev. B **54**, 11169 (1996).

[36]  G. Kresse and D. Joubert, Phys. Rev. B **59**, 1758 (1999).

[37]  J. Tersoff and D. R. Hamann, Phys. Rev. B **31**, 805 (1985).

[38]  H. J. Monkhorst and J. D. Pack, Phys. Rev. B **13**, 5188 (1976).

[39]  Y. Jiang, Y. Y. Sun, M. Chen, Y. Wang, Z. Li, C. Song, K. He, L. Wang, X. Chen, Q.-K. Xue, X. Ma, and S. B. Zhang, Phys. Rev. Lett. **108**, 066809 (2012).

[40]  P. Sessi, O. Storz, T. Bathon, S. Wilfert, K. A. Kokh, O. E. Tereshchenko, G. Bihlmayer, and M. Bode, Phys. Rev. B **93**, 035110 (2016).

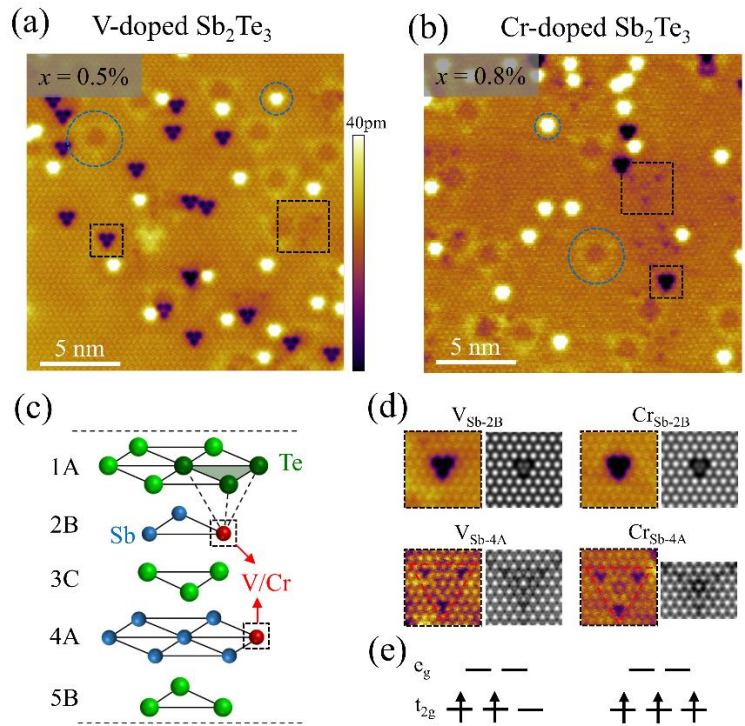

Fig. 1. Topographic image of (a) V-doped $Sb_2Te_3$ and (b) Cr-doped $Sb_2Te_3$ (Tunneling condition: $-1$ V, 0.5 nA). Black squares mark $V_{Sb}$ or $Cr_{Sb}$ defects at 2B or 4sA sites. Blue circles mark $Sb_{Te}$ defects at 1A and 5B sites. (c) Definition of atomic sites in the crystal structure of $Sb_2Te_3$. (d) STM images of individual magnetic defects. Colored images show experimental results; grey-scale images show the corresponding DFT simulations. (e) Crystal field splitting of $3d$ electrons of V and Cr.

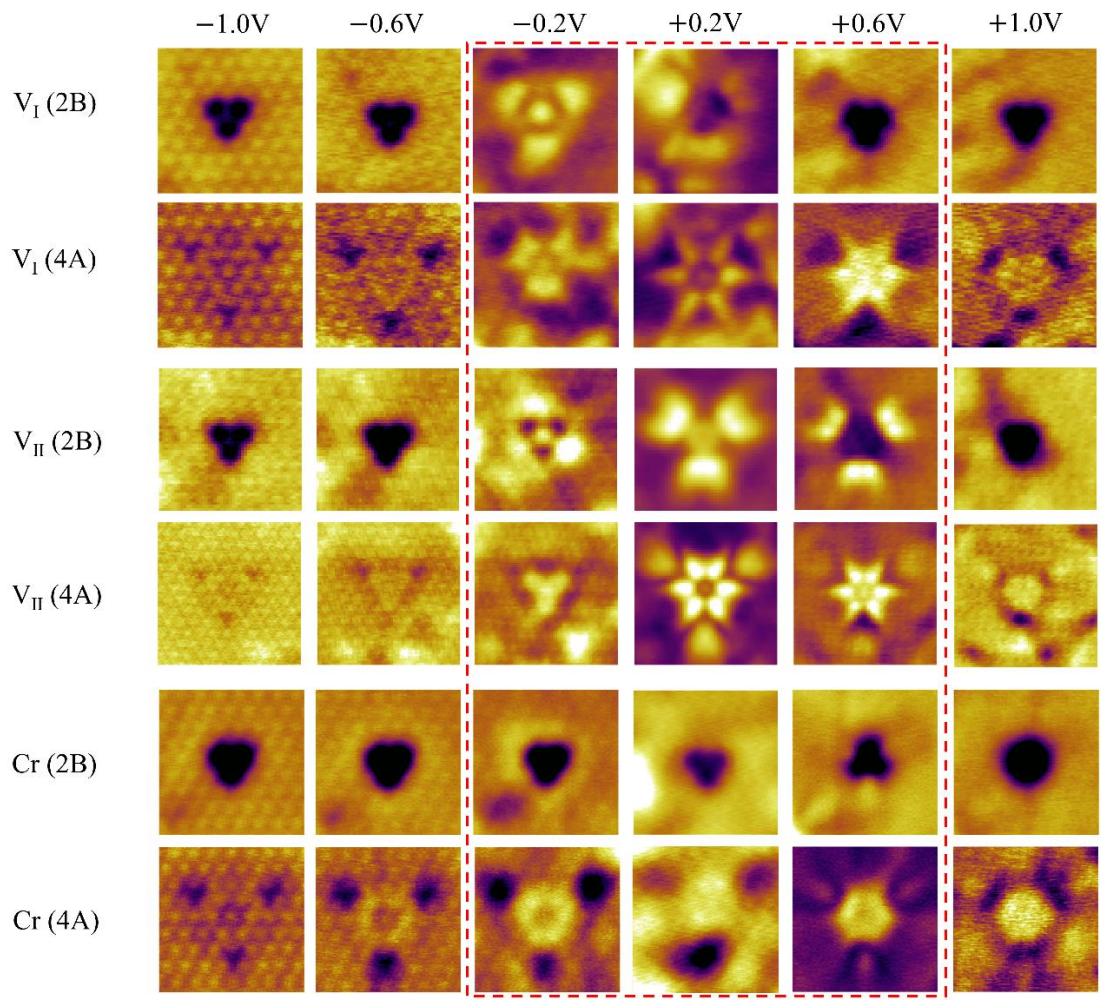

Fig. 2. Bias-dependent topographic images of V and Cr defects. The six rows show different types of defects: $V_I$ at 2B and 4A sites, $V_{II}$ at 2B and 4A sites, Cr at 2B and at 4A sites, respectively. The sample bias varies from −1.0 V to +1.0 V.

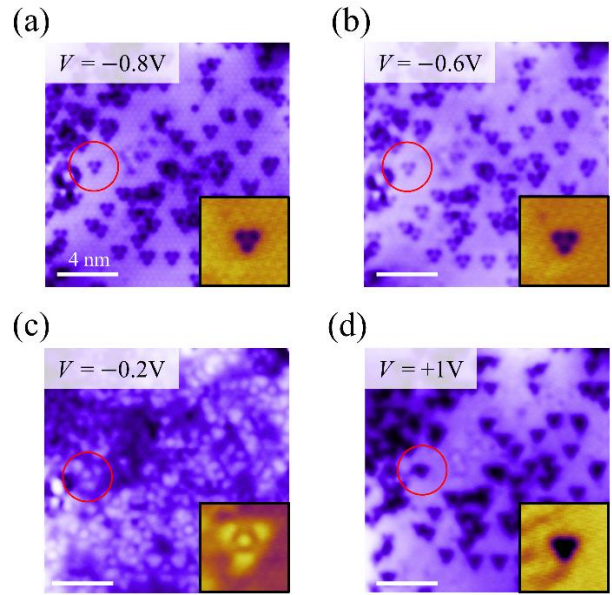

Fig. 3. Bias-dependent topographic images of thin-film $V_xSb_{2-x}Te_3$. A typical V defect at the 2B site is circled. Insets: defect image of $V_I$ in single crystal $V_xSb_{2-x}Te_3$.

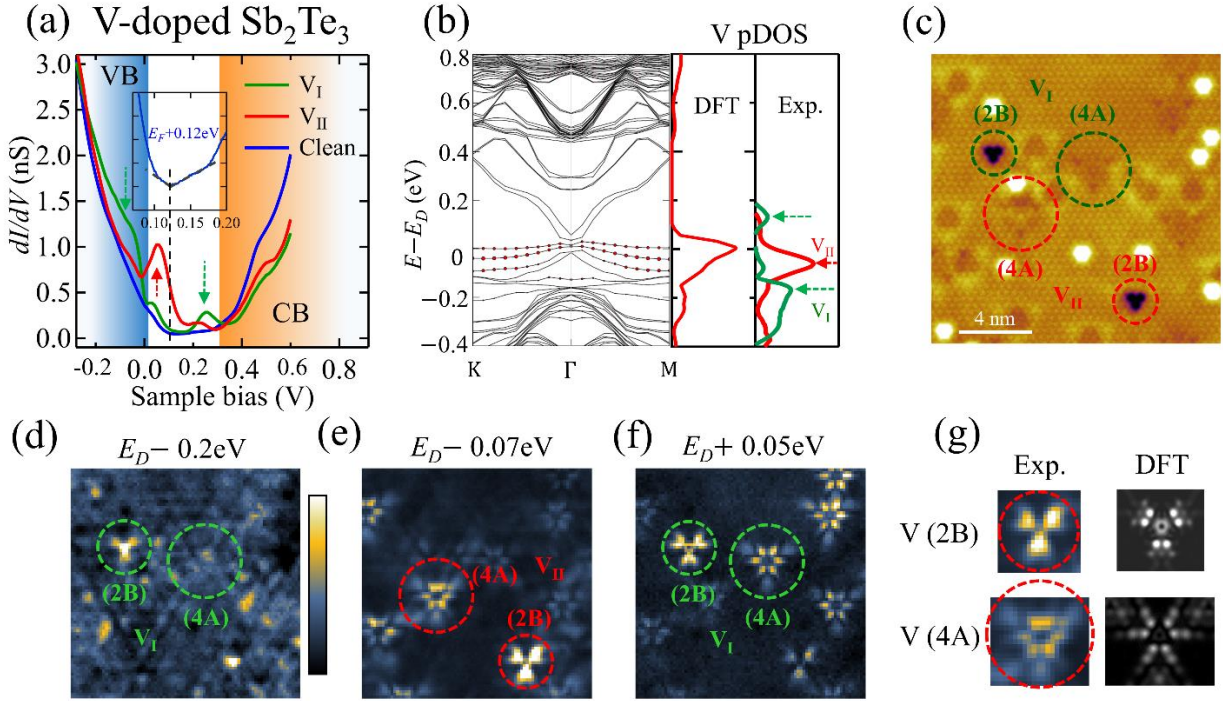

Fig. 4. Electronic defect states of V. (a) Average d$I$/d$V$ spectra in selective regions of V-doped Sb$_2$Te$_3$: green (red) on V$_I$ (V$_{II}$); blue in the defect-free area (−0.5 V, 1 nA). Inset: d$I$/d$V$ spectra of the surface states measured in defect-free areas. (b) DFT calculated band structure of V$_{0.03}$Sb$_{1.97}$Te$_3$ and partial DOS, as well as the d$I$/d$V$ difference between V sites with defect-free areas. The arrows indicate the energies of the defect states. (c) Topographic image of V-doped Sb$_2$Te$_3$ (−1.0 V, 0.5 nA), which contains both V$_I$ and V$_{II}$ at both 2B and 4A sites. (d-f) d$I$/d$V$ maps taken in the same area as (c) at $E_D$ − 0.2 eV, $E_D$ − 0.07 eV and $E_D$ + 0.05 eV, respectively. The defect states of V$_I$ are pronounced in (d) and (f), while those of V$_{II}$ in (e). (g) Zoom-in d$I$/d$V$ images of in-gap defect states of V$_{II}$ and the DFT simulations.

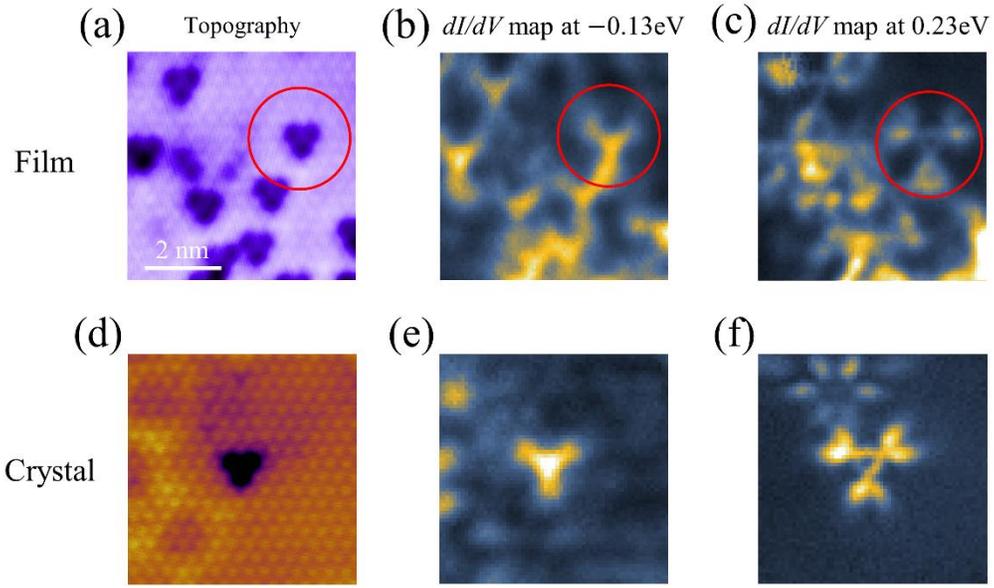

Fig. 5. The localized defect states in thin-film $V_x Sb_{2-x} Te_3$. (a) Topographic image ($-0.6$ V, 3 nA). (b) $dI/dV$ map at $E = -0.13$ eV, showing the localized defect states of V below VBM. (c) $dI/dV$ map at $E = 0.23$ eV, showing the in-gap defect states of V. (d-f) The corresponding defect states of V in single crystal $V_x Sb_{2-x} Te_3$, which coincides with thin-film results.

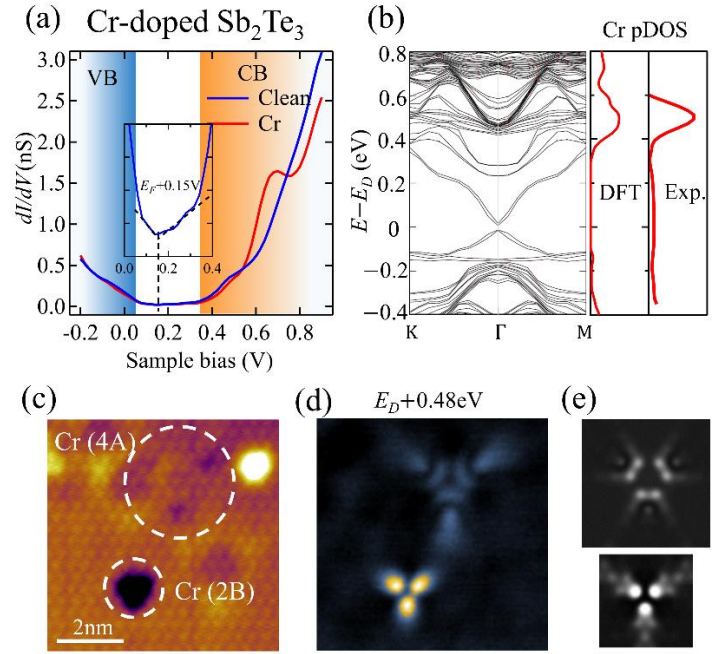

Fig. 6. Electronic defect states of Cr. (a) Average d$I$/d$V$ spectra in selective regions of Cr-doped Sb$_2$Te$_3$: red near Cr and blue in the defect-free area (+0.9 V, 1 nA). Inset: d$I$/d$V$ spectra of the surface states measured in defect-free areas. (b) DFT calculated band structure of Cr$_{0.03}$Sb$_{1.97}$Te$_3$ and the partial DOS of Cr, as well as the experimental d$I$/d$V$ difference between Cr sites with defect-free areas. (c) Topographic image of Cr-doped Sb$_2$Te$_3$ (−1.0 V, 0.5 nA), which contains Cr defects at both 2B and 4A sites. (d) d$I$/d$V$ map taken in the same area as (c) at $E_D$ + 0.48 eV, showing the local defect states high above the CBM. (e) the DFT simulated spatial distribution of defect states of Cr.

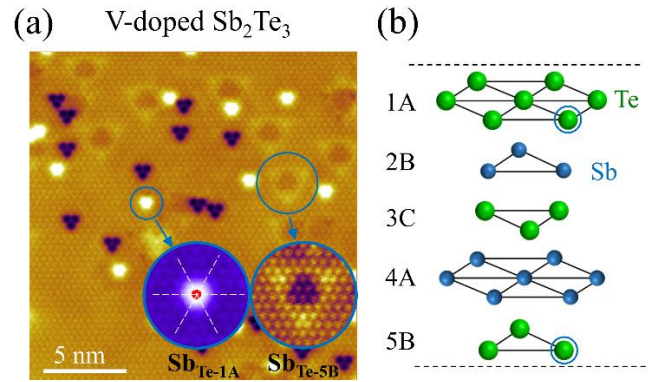

Fig. 7. Native point defects in V- or Cr- doped $Sb_2Te_3$. (a) Topographic image same as Fig 1(a). Blue circles mark the anti-site point defect $Sb_{Te}$ at 1B and 5A sites. The insets show the zoom-in STM images. (b) Illustration of the atomic position of $Sb_{Te}$ in the lattice.